\documentclass[12pt, a4paper, oneside]{article}
\usepackage{amsmath, amsthm, amssymb, appendix, bm, graphicx, hyperref, mathrsfs,natbib}

\newcommand{\aap}{    {\it Astron. Astrophys.}}

\newcommand{\apj}{    {\it Astrophys. J.}}

\newcommand{\cjaa}{   {\it Chin. J. Astron. Astrophys.}}

\newcommand{\pasj}{   {\it Pub. Astron. Soc. Japan}}

\newcommand{\solphys}{{\it Solar Phys.}}

\chardef\us=`\_
\begin{document}

\title{A Study on Correcting the Effect of Polarization Crosstalk in Full-Disk Solar Photospheric Magnetic Fields Observations}
\author{Liu, S.$^{1,2}$, Su, J.T.$^{1,2}$, Bai, X.Y$^{1,2}$., Deng, Y.Y.$^{1,2}$, Chen, J$^{1,2}$.,\\Song, Y.L.$^{1,2}$, Wang, X.F.$^{1,2}$, Xu, H.Q.$^{1,2}$, Yang, X.$^{1,2}$\\
$^{1}$ National Astronomical Observatories,\\ CAS, Beijing, 100101, China\\
$^{2}$  School of Astronomy and Space Sciences, University of CAS,\\ Beijing, 100101, China\\

lius@nao.cas.cn

}

\date{     }
\maketitle
\begin{abstract}
Magnetography using magnetic sensitive lines is regarded traditionally as the main instrument for measuring the magnetic field of the whole Sun.
Full polarized Stockes parameters ($I$, $Q$, $U$, $V$) observed can be used to deduce the magnetic field under specific theoretical model or inversion algorithms.
Due to various reasons, there are often cross-talk effects among Stokes signals observed directly by magnetographs.
Especially, the circular polarized signal $V$ usually affects the linear polarized ones $Q$ and $U$ seriously, which is one of the main errors of the value of the transverse magnetic field (parallel to the solar surface) that is related to $Q$ and $U$. The full-disk magnetograph onboard the Advanced Space based Solar Observatory (ASO-S/FMG) is designed to observe Stockes parameters to deduce the vector magnetic field. In this paper, the methods correcting the effects of cross-talk $V$ to $Q$ and $U$ are based on the assumption of perfectly symmetric Q and U and anti-symmetric Stokes V profiles and a new method to reduce the crosstalk effect under observation mode of FMG is developed. Through the test, it is found that the two methods have better effect in cross-talk removal in the sunspot region, and have better consistency. Addtionally, the developed methodcan be applied to remove the cross-talk effect using only one group of $Q$, $U$ and $V$ images observed at one wavelength position.
\end{abstract}

\section{Introduction}\label{S-Introduction} 

The measurement of the solar magnetic field is an important part in the study of solar activities.
The magnetic fields of active regions mostly contribute to the local energy release, such as solar flares, filament eruptions and small-scale activities.
The full-disk magnetic field observation is very necessary to understand magnetic field creation, large-scale magnetic connections, large-scale properties of magnetic field and coronal mass eruptions.
The full-disk magnetograph onboard the Advanced Space based Solar Observatory (ASO-S/FMG) is a new instrument to observe full-disk magnetic fields, which will surely contribute to studies of solar activity.
FMG uses a magnetic sensitive line of Fe I $\lambda$5234.19~\AA~to observe full-disk magnetic field. FMG is designed to be a filter type instrument to obtain full polarized Stockes parameters ($I$, $Q$, $U$, $V$), which are used to calibrate magnetic field on the solar surface. During the operation of FMG, a few limited spectral positions inside the line are selected to observe Stockes parameters directly.

Many observation factors (the accuracy of wavelength position, wavelength shift due to filter properties or solar rotation, magnetic saturation, scattered light and so on) can affect the accuracy of magnetic field measurements,
while for polarimetric instruments, the quality of Stockes parameters affect most directly the accuracy of the magnetic field derived. As a result, Stokes parameters with high-accuracy are the first step to obtain accurate magnetic fields.
However, there is a very difficult problem to solve that there exist cross-talk effects among Stokes parameters, which means that individual Stokes signals can be contaminated by the others more or less (\citeauthor{1983SoPh...88...51W}~\citeyear{1983SoPh...88...51W}; \citeauthor{1992SPIE.1746..281W}~\citeyear{1992SPIE.1746..281W}; \citeauthor{2007ApJ...666..559S}~\citeyear{2007ApJ...666..559S}).

Horizontal magnetic fields with a field strength above the equipartition value (400 G) in the solar photosphere are unstable, which then be pushed by the force of buoyancy upright with a vertical orientation relative to the solar surface (\citeauthor{1979SoPh...62...15S},~\citeyear{1979SoPh...62...15S}). 
The lateral expansion of magnetic flux concentrations (e.g. magnetic elements and sunspots) can only be set in the upper photosphere and chromosphere, where the density of gas has dropped sufficiently (\citeauthor{1986A&A...170..126S},~\citeyear{1986A&A...170..126S}).
As result, a dominance of vertical magnetic fields in the solar photosphere, where the current measurements are taken. The amplitude of the Stokes parameters QU and V has a different dependence on the inclination $\gamma$ between the line of sight (LOS) and the magnetic field orientation. The amplitude of circular polarization Stokes $V$ is proportional to cos($\gamma$), while the amplitudes of linear polarization $Q$ and $U$ are proportional to sin$^2$($\gamma$) (\citeauthor{1989ApJ...343..920J}, \citeyear{1989ApJ...343..920J}).
Because of the LOS over most of the disk and the magnetic field orientation are close to the solar vertical, the observed Stokes V amplitudes are usually much larger than those of Stokes Q or U on most locations.  Consequently, once circular polarization $V$ contaminates the linear ones $Q$ and $U$, the influence of $V$ to $Q$ and $U$ would become serious and the most important one to be corrected urgently.

The inevitable cross-talk effects among Stokes parameters $Q$, $U$ and $V$ can not be removed completely. The standard method for removal of cross-talk is a polarimetric calibration of the telescope and instrument optics (\citeauthor{1997ApJS..110..357S}, \citeyear{1997ApJS..110..357S}; \citeauthor{2005A&A...437.1159B}, \citeyear{2005A&A...437.1159B}; \citeauthor{2005A&A...443.1047B}, \citeyear{2005A&A...443.1047B}; \citeauthor{2011A&A...531A...2S}, \citeyear{2011A&A...531A...2S}; \citeauthor{2017JATIS...3a8002H}, \citeyear{2017JATIS...3a8002H};). 
The term $cross-talk$ used in this paper is only used for the removal of residual cross-talk (i.e., the residuals left over after the regular polarimetric calibration). 
The methods mentioned here are based on the assumption that V$_{obs}$~$\approx$~V$_{true}$, which usually needs to be reached by a polarimetric calibration before.
As for the removal of these residual cross-talk after polarimetric calibration, especially the cross-talk $V$ to $Q$ and $U$, which is the most important one, some effective methods or algorithms are proposed occasionally, especially for new developed instruments. \citeauthor{1992SPIE.1746..281W}~(\citeyear{1992SPIE.1746..281W}) developed and verified a method to correct crosstalk of circular-to-linear $V-Q/U$, under the assumption that $Q$ and $U$ profiles are completely symmetric and the asymmetries are results from errors of instruments. This kind of method has been used and tested in active region magnetic field measurements ( \citeauthor{1985SoPh...97..223M}, \citeyear{1985SoPh...97..223M}; \citeauthor{1992SPIE.1746..281W}, \citeyear{1992SPIE.1746..281W}). \citeauthor{2007ApJ...666..559S}~(\citeyear{2007ApJ...666..559S}) extended the application of this method to full-disk magnetic field observation. \citeauthor{2007ApJ...666..559S}~(\citeyear{2007ApJ...666..559S}) employ directly observations of two special spectral position, where $V$ profile is approximately symmetric around the position of their center. This symmetrical observation in pairs is also based on the symmetry and asymmetry of spectral profile in theory for circular and linear polarization signals, respectively.

In this paper, we focus on the removal of cross-talk from V to Q/U, which affects the accuracy of transverse field mainly, with the purpose to develop and test methods correcting crosstalk effect for the Full-disk solar vector MagnetoGraph (FMG) onboard the Advanced Space-based Solar Observatory (ASO-S).
The followings parts are observations in sections~\ref{sec:obser}, the descriptions of methods in section~\ref{sec:method}, the results with error analysis in section~\ref{sec:results}, the discussions in section~\ref{sec:disc}, \textbf{whereas the} last section~\ref{sec:conc} gives the conclusions.

\section{Observations}\label{sec:obser}
\begin{figure}
 \centerline{\includegraphics[width=1\textwidth,clip=]
{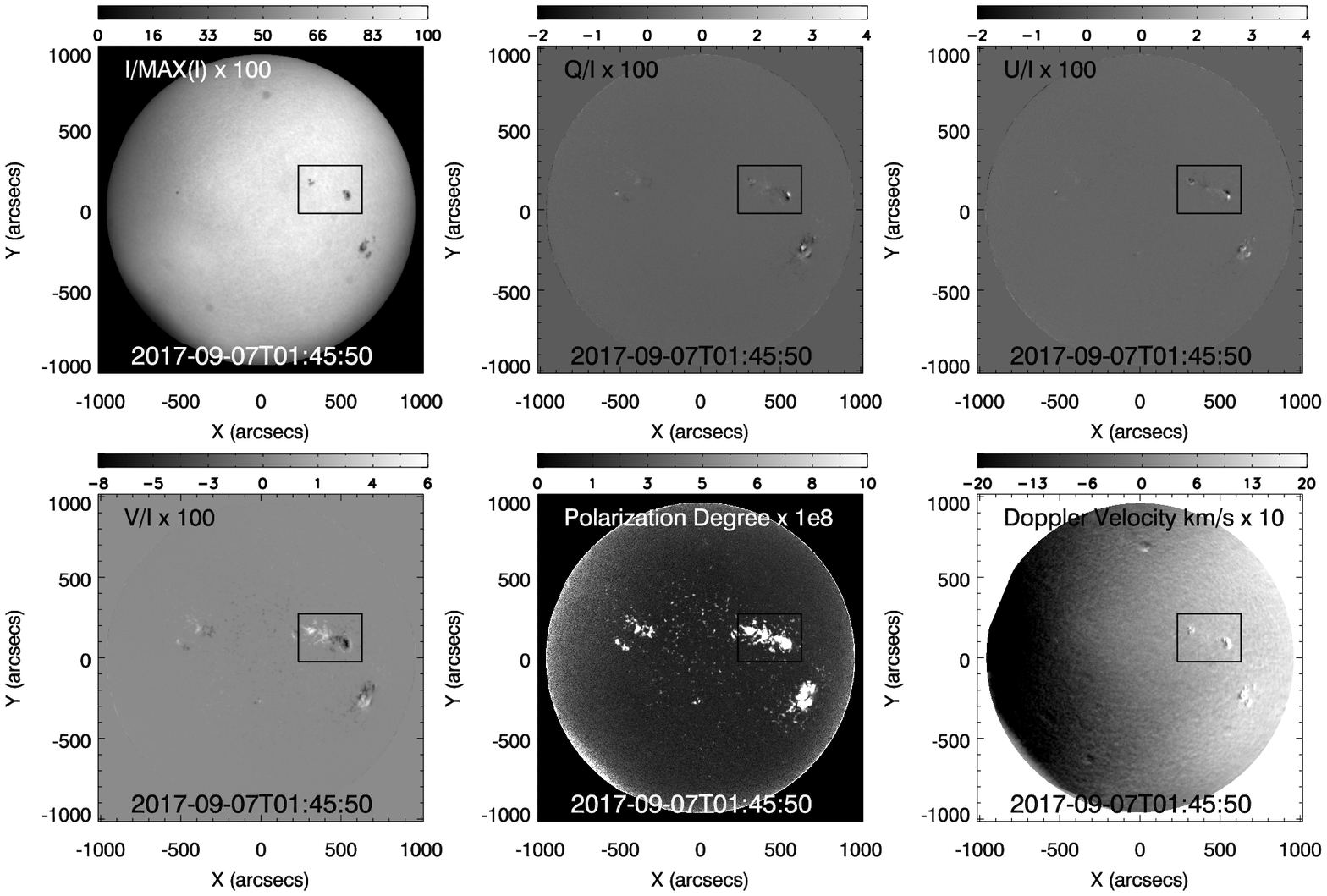}}
   \caption{The SMAT full-disk observations of $I$/max($I$), $Q$/$I$, $U$/$I$, $V$/$I$, polarization degree and Doppler velocity images, the black rectangle is the region magnified in Figure \ref{iquvpdvelo_local}.} \label{iquvpdvelo}
\end{figure}

\begin{figure}
 \centerline{\includegraphics[width=1\textwidth,clip=]
{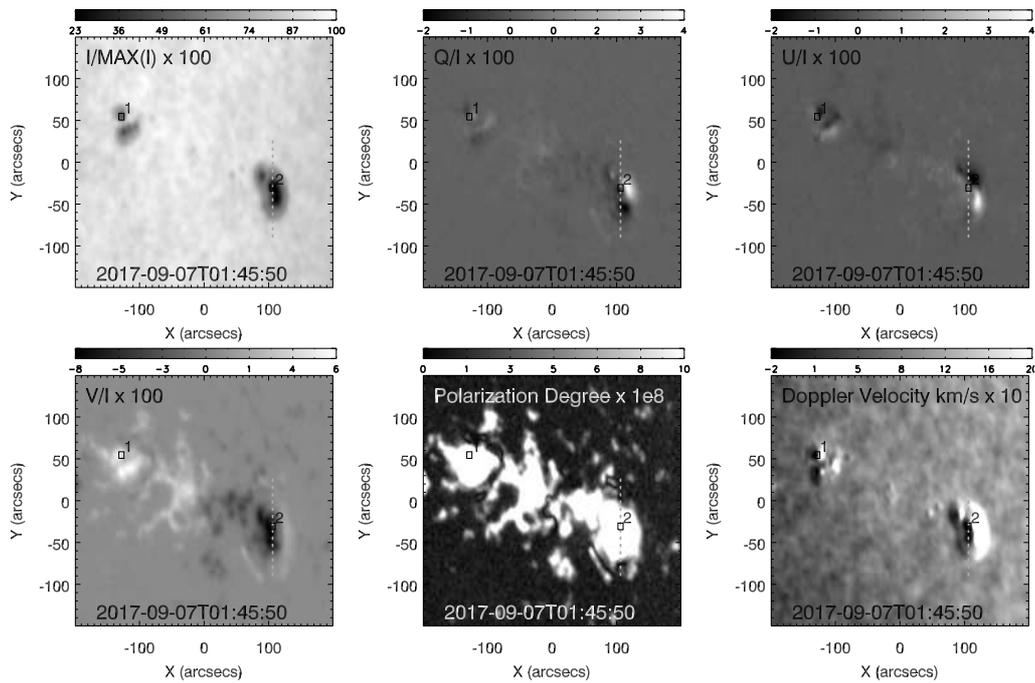}}
   \caption{The SMAT cutout images from Figure \ref{iquvpdvelo} indicated by black rectangle, the black squares 1 and 2 are selected position to produce spectrum in Figures \ref{profile_plot} and \ref{profile_plot_beforeaftercorr_method12}, while the white dotted line is the slit to create spectrum in Figure \ref{profile_plot_slit_beforeaftercorr_method12}.} \label{iquvpdvelo_local}
\end{figure}

\begin{figure}
 \centerline{\includegraphics[width=1\textwidth,clip=]
{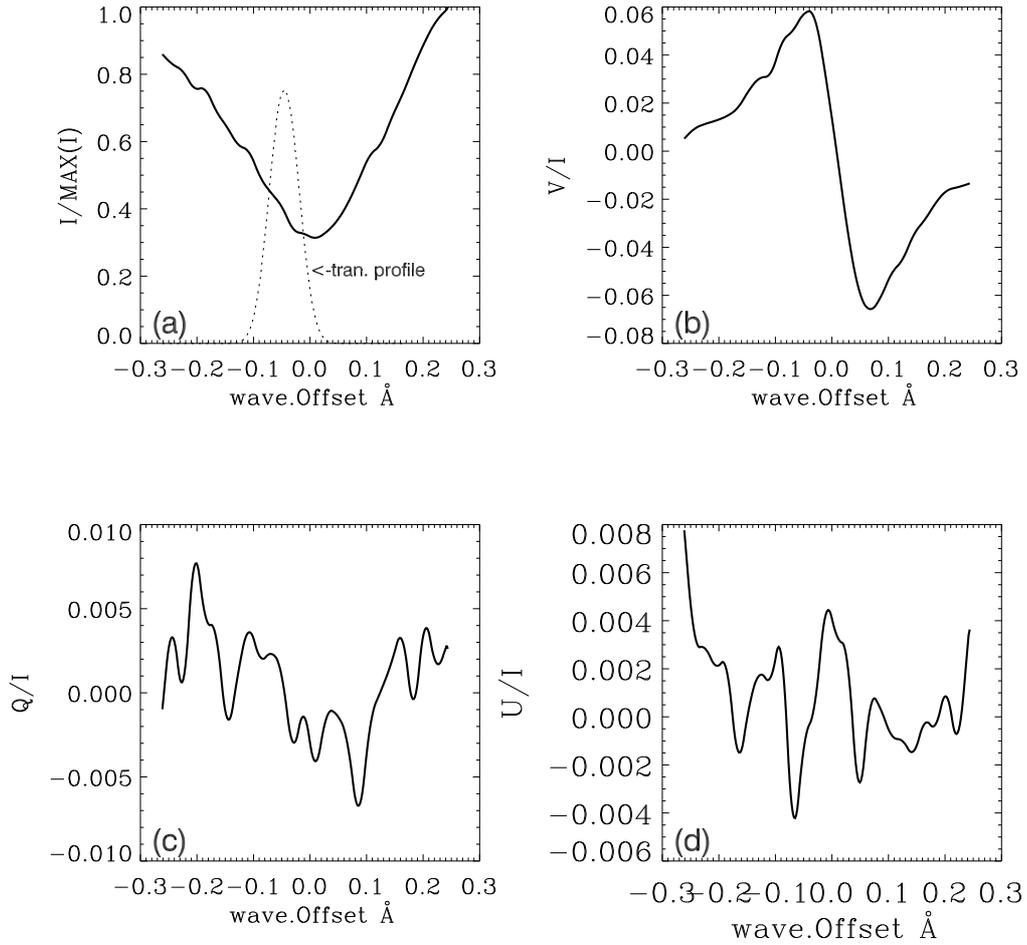}}
   \caption{The SMAT normalized $I$/max($I$), $Q$/$I$, $U$/$I$, $V$/$I$ spectra from a position labelled by the small black squre 1 in Figure \ref{iquvpdvelo_local}, the transmission profile of the Lyot filter using green dashed line is overplotted in panel (a).} \label{profile_plot}
\end{figure}

FMG (\citeauthor{2019RAA....19..157D}, \citeyear{2019RAA....19..157D}) uses filters of the same type and the spectral line of Fe I $\lambda$5234.19 \AA~as the Solar Magnetism and Activity Telescope at Huairou Solar Observing
Station  (SMAT/HSOS: \citeauthor{2007ChJAA...7..281Z}, \citeyear{2007ChJAA...7..281Z};  \citeauthor{2007ApJ...666..559S}, \citeyear{2007ApJ...666..559S}; \citeauthor{lin2013}, \citeyear{lin2013};  \citeauthor{2019RAA....19..157D}, \citeyear{2019RAA....19..157D}; \citeauthor{2019RAA....19..156G}, \citeyear{2019RAA....19..156G}; \citeauthor{2019RAA....19..161S}, \citeyear{2019RAA....19..161S})
to observe the full-disk magnetic field, so the observations obtained from SMAT with scanning mode are used as experimental data to verify the effectiveness of the methods to remove cross-talk effects. The dataset used here was observed on 2017-09-07 from 01:33:05 to 02:12:09 UT, the scaning mode set: -0.30 \AA~to 0.30 \AA~with steps of 0.02 \AA, $\sim$9~s exposure time with 256 frames. The observed $I$, $Q$, $U$ and $V$ images are  992 $\times$ 992 with the pixel size of 1.983 arcsec/pixel. Figure \ref{iquvpdvelo} shows the $I$/max($I$), $Q$/$I$, $U$/$I$, $V$/$I$ (hereafter  $Q$, $U$, $V$ indicates $Q$/$I$, $U$/$I$, $V$/$I$ if there is no special points), polarization degree and Doppler velocity images of full-disk observations. For Doppler velocity map, it is used for modification of real spectral position. Figure \ref{iquvpdvelo_local} shows the local enlarged image indicated by the rectangle labelled in Figure \ref{iquvpdvelo} with strong polarization signals. To check the properties of observations, rms noise level in  $Q$/$I$, $U$/$I$, $V$/$I$ are calculated at quiet region without significant polarization signal, which are $1 \times 10^{-4}$, $1 \times 10^{-4}$, $3 \times 10^{-4}$ for $Q/I$, $U/I$ and $V/I$ for integration of all wave observations, respectively. While they are  $7 \times 10^{-4}$, $7 \times 10^{-4}$, $15 \times 10^{-4}$ for $Q/I$, $U/I$ and $V/I$ at one fixed wavelength location. Figure \ref{profile_plot} shows an example of normalized  $I$/max($I$), $Q$/$I$, $U$/$I$, $V$/$I$ spectra from a location labelled by the small squre 1 in Figure \ref{iquvpdvelo_local} with significant polarization signals. In panel a, the transmission profile of the Lyot filter using green dashed line is overplotted to compare spectral bandpass with the observed line width. The passband of the spectral line of Fe I $\lambda$5234.19 \AA~is about 0.125 \AA~ and the observed line width is about 0.3 \AA, which can also refer to the paper of \citeauthor{2019RAA....19..157D} (\citeyear{2019RAA....19..157D}) and \citeauthor{2007ChJAA...7..281Z} (\citeyear{2007ChJAA...7..281Z}). 

A group of quasi pure $I$, $Q$, $U$ and $V$ images observed taken by the Near Infra-Red Imaging Spectropolarimeter (NIRIS: \citeauthor{2012ASPC..463..291C}, \citeyear{2012ASPC..463..291C}) installed on Goode Solar Telescope (GST: \citeauthor{2012ASPC..463..357G}, \citeyear{2012ASPC..463..357G}) at Big Bear Solar Observatory (BBSO) are used in this paper. NIRIS can obtain photospheric magnetic fields by spectropolarimetric observations of the Fe I 1.56 $\mu_{m}$ line (0.25 \AA~bandpass). 
$I$/max($I$), $Q$/$I$, $U$/$I$, $V$/$I$ images of NOAA 12600 used here are observed on 2016-10-15 from 19:18:53 to 19:1925 UT, the spectra is set from -3.0 \AA~to 3.0 \AA~ with \textbf{a} step of 0.11 \AA. The image is 720 $\times$ 720 with the pixel size of 0.0244 arcsec/pixel, giving a total field of view of 17.5 $\times$ 17.5 arcsec.
Figure \ref{NIRIS} shows $I$/max($I$), $Q$/$I$, $U$/$I$, $V$/$I$ images observed by NIRIS, and the correlations between  $V$ and $Q$/$U$ are calcuated and labeled with the values of $1.09 \times 10^{-4}$ and $1.25 \times 10^{-4}$, respectively. 
It is noted that polarization signal in the umbra is lower than in the penumbra, this maybe due to polarization intensity characteristics ('magnetic saturation'), when image is obtained at one wavelength position.
For this dataset, the rms noise level are $3 \times 10^{-4}$, $3 \times 10^{-4}$, $1 \times 10^{-4}$ for $Q/I$, $U/I$ and $V/I$ for integration of all wave observation, respectively, while they are $10 \times 10^{-4}$, $9 \times 10^{-4}$, $6 \times 10^{-4}$ for $Q/I$, $U/I$ and $V/I$ at one fixed wavelength location. Figure \ref{profile_plot_niris} shows an example of normalized $I$/max($I$), $Q$/$I$, $U$/$I$, $V$/$I$ spectra from a location labelled by the small square in Figure \ref{NIRIS} with significant polarization signal. On the whole these patterns of spectra are normal ones, since the spectra of $V$ keep good symmetry property, while $Q$ and $U$ has good antisymmetry property basically.

\section{Methods}\label{sec:method}
\subsection{Method I: symmetry properties of Stokes QUV spectra}
The method of symmetry properties of Stokes QUV spectra (referee as Method I) is extended from  \citeauthor{1992SPIE.1746..281W}~(\citeyear{1992SPIE.1746..281W}) and  \citeauthor{2007ApJ...666..559S}~(\citeyear{2007ApJ...666..559S}).
The method developed by \citeauthor{2007ApJ...666..559S}~(\citeyear{2007ApJ...666..559S}) is focused on correcting the effects of cross talk $V$ to $Q$ and $U$ for full-disk magnetic field obtain by HSOS/SMAT (\citeauthor{2007ChJAA...7..281Z}, \citeyear{2007ChJAA...7..281Z})). Hence, following \citeauthor{2007ApJ...666..559S}~(\citeyear{2007ApJ...666..559S}), the cross-talk coefficients for symmetrical Stokes $Q$ and $U$ and anti-symmetrical Stokes $V$ signals can be derived through 
\begin{equation}
\begin{array}{ll}
Q^{'}_{\lambda_{0}+\delta\lambda_{1}}-Q^{'}_{-\lambda_{0}+\delta\lambda_{2}}=C_{q}(x,y)(V_{\lambda_{0}+\delta\lambda_{1}}-V_{-\lambda_{0}+\delta\lambda_{2}})
\end{array}
\end{equation}
\begin{equation}
\begin{array}{ll}
U^{'}_{\lambda_{0}+\delta\lambda_{1}}-U^{'}_{-\lambda_{0}+\delta\lambda_{2}}=C_{u}(x,y)(V_{\lambda_{0}+\delta\lambda_{1}}-V_{-\lambda_{0}+\delta\lambda_{2}})
\end{array}
\end{equation}
Then the corrected  $Q^0$ and $U^0$ can be calculated by
\begin{equation}\label{correct1}
\begin{array}{ll}
Q^{0}=Q^{'}-C_{q}V, U^{0}=U^{'}-C_{u}V,
\end{array}
\end{equation}

Here $Q^0$ and $U^0$ are actual linear polarization signals, $Q^{'}$ and $U^{'}$ as the contaminated ones.
$C_{q}(x,y)$ and $C_{u}(x,y)$ are defined as cross-talk coefficients, which indicate the fractions 
$V$ crosstalk in $Q^0$ and $U^0$ signals, respectively.
The subscript $\lambda_{0}$ indicates the actual filter position and  $\delta\lambda_{1}$ and  $\delta\lambda_{2}$ refer to the wavelength shifts relative to $\lambda_{0}$ and $-\lambda_{0}$.
During observation for the full-disk observations, $\delta\lambda_{1}$ =  $\delta\lambda_{2}$ are required. It is noted that for this method, the assumption requires to know true $V$ signals, but the only quantity $V^{'}$ ($V$ contaminated by others Stokes parameters) is available, $V$ and $V^{'}$ is not the same whenever there are effects of cross-talk. Fortunately, the contamination from $Q$ and $U$ to $V$ is small, while the contamination from $I$ to $V$ can be removed relatively easily.

\subsection{Method II: Minimal correlation of spatial QUV maps}
\begin{figure}
 \centerline{\includegraphics[width=1\textwidth,clip=]
{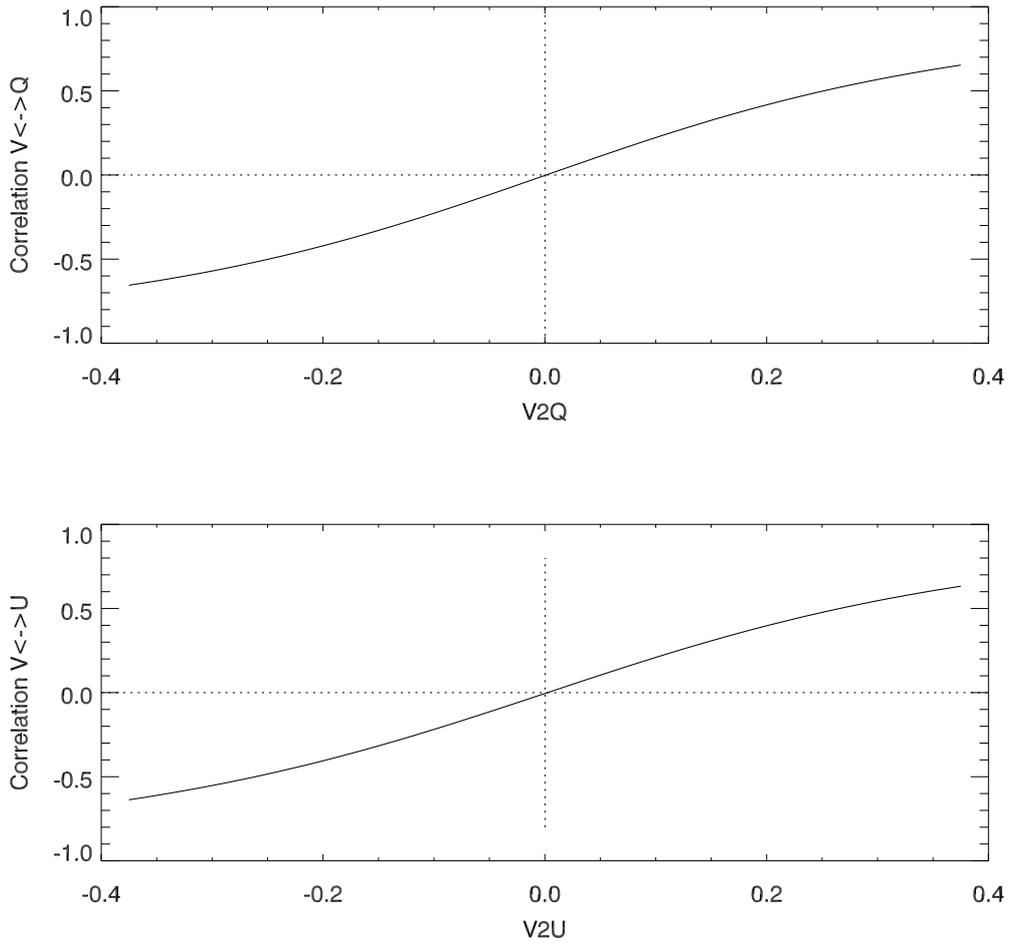}}
   \caption{The correlation coefficient between NIRIS maps of $V$ and $Q$ (top) and $U$ as a function of the $V->QU$ cross-talk set. The vertical dashed lines indicate zero cross-talk. The minimal correlation is reached at $1 \times 10^{-4}$ and $1 \times 10^{-4}$ for $Q$ and $U$, respectively.} \label{corrvscrossniris}
\end{figure}

\begin{figure}
 \centerline{\includegraphics[width=1\textwidth,clip=]
{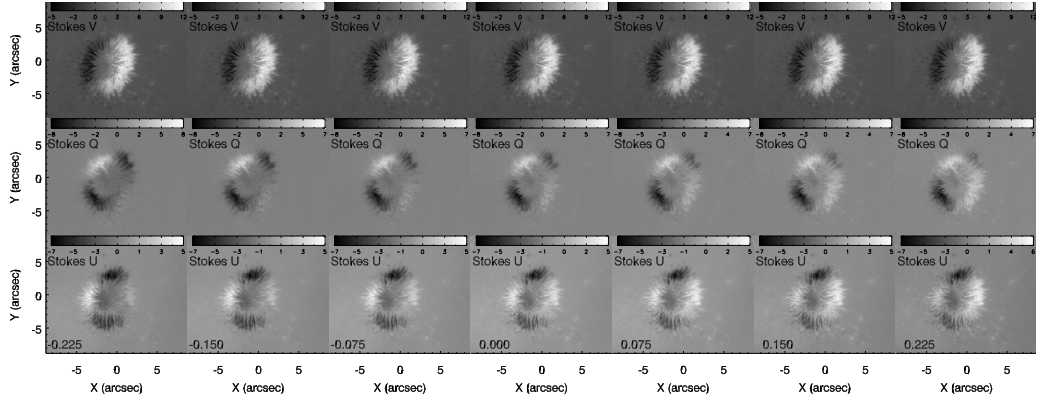}}
   \caption{The NIRIS maps of $V$ (top row) and $Q$ (middle row) and $U$ (bottom row) at diﬀerent values of the $V→QU$ cross-talk set. The cross-talk value is indicated at the bottom of the $QU$ panels.} \label{corrvscrossnirisimage}
\end{figure}

The method of minimal correlation of spatial QUV maps (referee as Method II) is under on the assumption that $QU$ and $V$ have low or zero correlation in their spatial distribution, which is based on the physics that the vertical central fields with horizontal canopies (the expansion of the flux concentration and makes increasingly more inclined magnetic field lines with increasing height) and the different dependence of $QU$ and $V$ on inclination $\gamma$ ($V$ $\propto$ cos($\gamma$), $Q$ and $U$  $\propto$ sin$^2$($\gamma$)). The spatial pattern of any type of magnetic field concentration is a central maximum of the $V$ signal with large inclination that decreases with
the radial distance, while $Q$ and $U$ have a minimum at center with an increase in the radial direction. $Q$ and $U $ also show an azimuthal variation with neutral lines and iterative maxima, although the LOSs and more complex magnetic topologies can modulate that pattern. Additionally, due to the dependence of $QU$ and $V$ on inclination $\gamma$, the amplitudes of $Q$ and $U$ are usually lower than for those of $V$. All above facts means that minimal correlation between Stokes $V$ and $Q$/$U$ should be reached in absence of cross-talk, which can be seen both in observation (e.g. active region vector magnetic field observed by solar Magnetic Field Telescope (SMFT) installed at HSOS, and the method was used many times) and in simulations. The NIRIS with quasi pure $I$, $Q$, $U$ and $V$ can be used to test the extent of this assumption (minimal correlation of spatial $QUV$ maps in absence of cross-talk). Figure \ref{corrvscrossniris} shows the correlation coefficient for the $QU$ adjusted by artificially added $C_{qu}$ and $V$ original one in maps of Figure \ref{NIRIS}. The minimal correlation is reached at about $1 \times 10^{-4}$ and $1 \times 10^{-4}$ for $Q$ and $U$, which is the same as the observations of NIRIS after cross-talk removal by standard polarimetric calibration ($1.09 \times 10^{-4}$ and $1.25 \times 10^{-4}$).
Figure \ref{corrvscrossnirisimage} shows images of $QU$ adjusted by different values of C$_{qu}$ and original $V$. The effect of the additional cross-talk introduced into $Q$/$U$ is that the pattern of $V$ become more evident as the increased values of C$_{qu}$. Based on the Figures \ref{corrvscrossniris} and  \ref{corrvscrossnirisimage}, it can be found that the method of minimal correlation of spatial QUV maps works for NIRIS cross-talk removal, hence the method can be applied to FMG tentatively.

Based on the above assumptions and tests, we develop a method to extract the signals of $V$ from the contaminated $Q$ and $U$ using image information for FMG full-disk observation.
Firstly, a small region is cut out from full-disk image for $V$, $Q$ and $U$; then the a series of set proportions of $V$ signal is subtract from $Q$/$U$ for this small local image; thirdly, we find the set proportion that can give the minimum of correlation between $Q$/$U$ and $V$, and this proportion is regarded as cross-talk coefficient of the center of that small image (not the whole small image, only is the pixel position of its center); at last the whole distribution of cross-talk coefficient can be obtain by travelling throughout the whole solar surface using small local region.

\section{Results}\label{sec:results}
\begin{figure}
   \centerline{\includegraphics[width=1\textwidth,clip=]{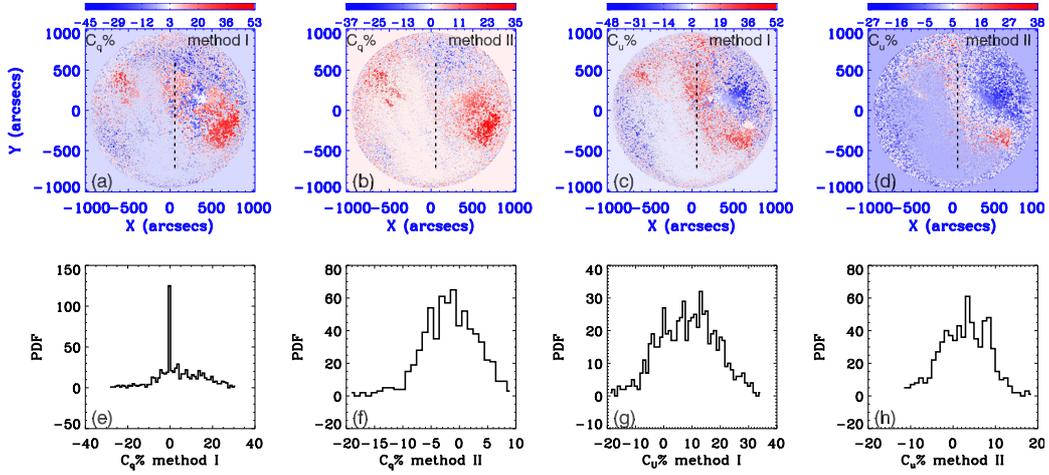}}
   \caption{The distributions of crosstalk coefficents of $C_{q}(x,y)$ and $C_{u}(x,y)$ with the restrictions with polarization degree $>$ 3 $\sigma$ on solar full-disk for method I and II labelled, the histogram plots are calculated from the black slits on full-disk for $C_{q}(x,y)$ and $C_{u}(x,y)$ obtained by method I and II, respectively.} \label{cqcumethod1and2}
\end{figure}

\begin{figure}
   \centerline{\includegraphics[width=1\textwidth,clip=]{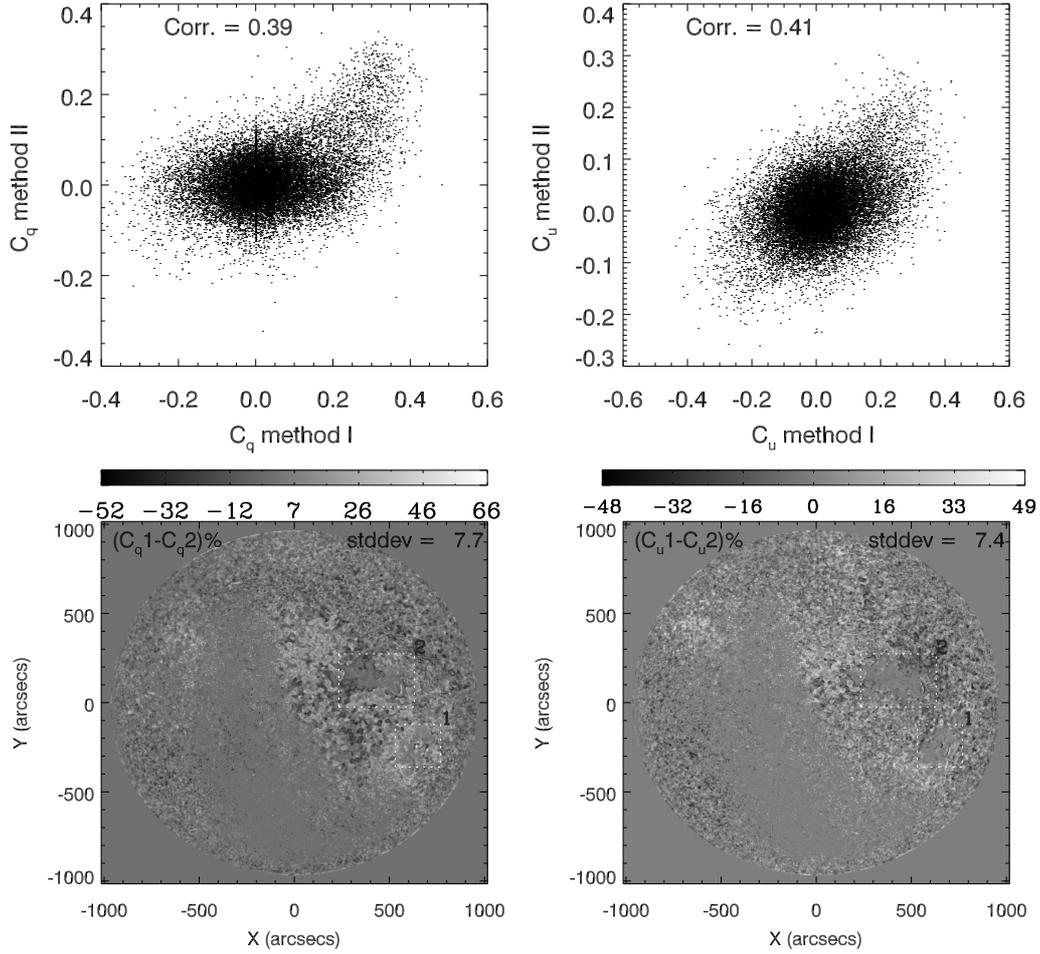}}
   \caption{Top row : the scatter-plots of crosstalk coefficents of $C_{q}(x,y)$ and $C_{u}(x,y)$ labelled on solar full-disk for method I and II, where the Corr. indicate the correlation of  $C_{q}(x,y)$ and $C_{u}(x,y)$ obtained from method I and II. Bottom row: the differences between $C_{q}1$ and $C_{q}2$ ($C_{u}1$ and $C_{u}2$), where $C_{q}1$ and $C_{u}1$ ($C_{q}2$ and $C_{u}2$) are obtaineb by method I and II, two white dotted rectangles 1 and 2 indicate the regions with significant polarization signals.} \label{method1vsmethod2}
\end{figure}

\begin{figure}
   \centerline{\includegraphics[width=1\textwidth,clip=]{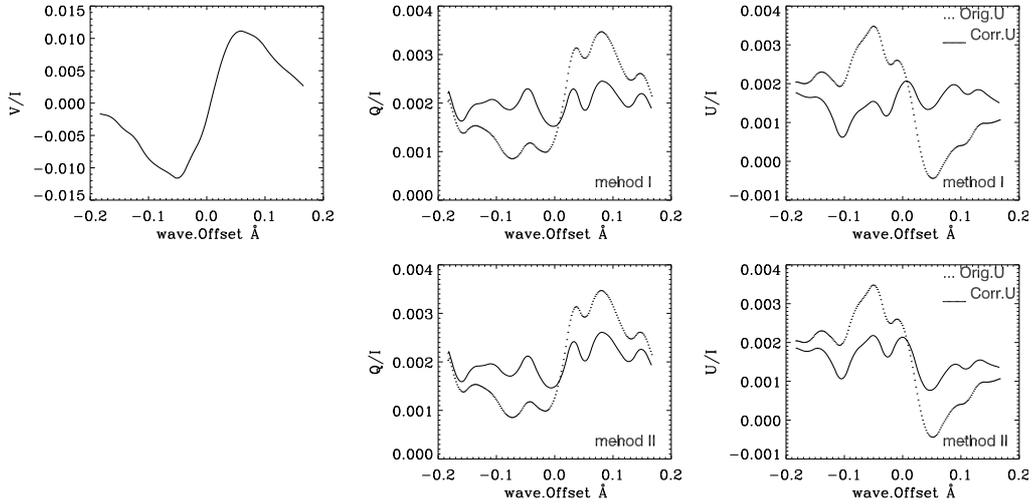}}
   \caption{The spectra of the point of $V$ and $Q$/$U$ before and after correction by method I and II, the points are  labelled in Figure \ref{iquvpdvelo_local} by the black square 2. The range of offset is $\pm$ 0.2\AA~with the line center. } \label{profile_plot_beforeaftercorr_method12}
\end{figure}

\begin{figure}
   \centerline{\includegraphics[width=1\textwidth,clip=]{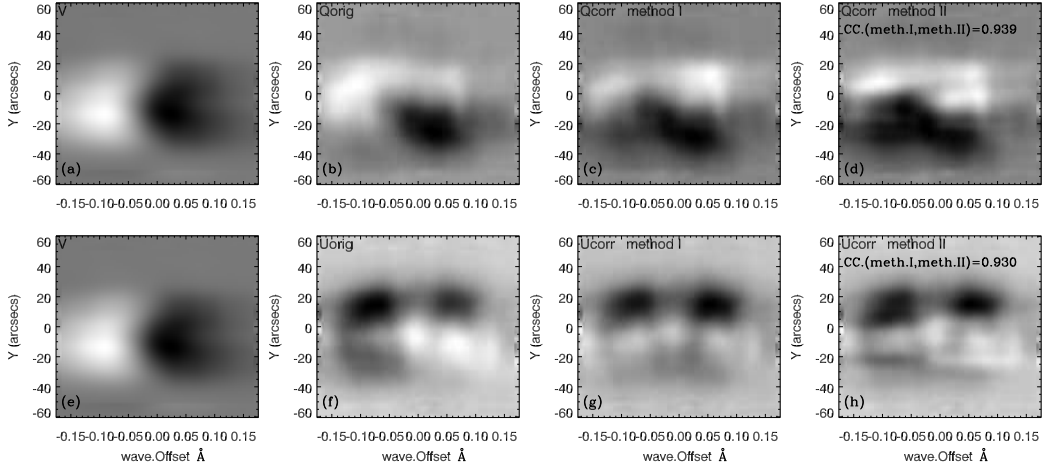}}
   \caption{The spectra of a slit of $V$ and $Q$/$U$ before and after correction by method I and II, the slits are labelled in Figure \ref{iquvpdvelo_local} by the whie dotted lines, and the correlations of images ((c) and (d), (g) and (h)) after corrections by method I and II are printed in panels (d) and (h).} \label{profile_plot_slit_beforeaftercorr_method12}
\end{figure}

Figure \ref{cqcumethod1and2} shows the distributions of cross-talk coefficents of $C_{q}(x,y)$ and $C_{u}(x,y)$ on solar full-disk obtained by method I and II labelled in panels (a), (b), (c) and (d), while the histogram plots in panels (e), (f), (g) and (h) are calculated from the black slits on full-disk for $C_{q}(x,y)$ and $C_{u}(x,y)$ by method I and II. It can be found that the distributions of cross-talk coefficents of $C_{q}$/$C_{u}$ have the similar patterns for method I and II (here the restrictions is added to the locations with polarization degree $>$ 3 $\sigma$, $\sigma$ is the stddev of polarization degree at the quiet region), however there should exist some differences between two methods that can be seen from histgram plots in panels (e), (f), (g) and (h). The top row of Figure \ref{method1vsmethod2} shows the scatter-plots of crosstalk coefficents of $C_{q}$ and $C_{u}$ labelled on solar full-disk for both of method I and II, at the mean time the correlations of $C_{q}$ and $C_{u}$ calculated by method I and II are labelled in each panels. It is found that the correlations of $C_{q}(x,y)$ and $C_{u}(x,y)$ calculated by methodI and II is about 40\% on the whole. The bottom row in Figure \ref{method1vsmethod2} shows the differences between $C_{q}1$ and $C_{q}2$ ($C_{u}1$ and $C_{u}2$), where subscript 1 and 2 means method I and II. where the stddev of (C$_{q}1$-$C_{q}2$) and (C$_{u}1$-$C_{u}2$) calculated are 0.077 and 0.074, respectively, it can be found that the deviations between $C_{q}1$ and $C_{q}2$ ($C_{u}1$ and $C_{u}2$) of the region with strong polarization are relatively small, which can be see from the regions with significant polarization signals labelled by two white dotted rectangles 1 and 2. Then, an example of the images of $Q$ and $U$ observed at -0.12\AA~before and after cross-talk corrected by cross-talk coefficents of $C_{q}(x,y)$ and $C_{u}(x,y)$ in Figure \ref{cqcumethod1and2} are shown in Figure \ref{cqcucorrectmethod1and2}. In this Figure, the white rectangles highlight the regions with evident polarization signal. The signals of $V$ are weakened in image of $Q$ and $U$ after cross-talk removal through comparisons of highlight region in detail, which can be seen more clearly in the magnification images in Figure \ref{cqcucorrectmethod1and2_magnif}. In Figure \ref{cqcucorrectmethod1and2_magnif}, the magnification images of the cut region labelled by rectangle 1 in Figure \ref{cqcucorrectmethod1and2} are highlighted, additionally the image of $V$ and polarization degree are appended in the left column, and the correlations of  $V$ and $Q$/$U$ before and after correction are printed in each panel individually. It is found that the contaminations of $V$ in $Q$ and $U$ are weakened obviously after cross-talk effects removal, which can also be seen from the lower values of correlations between $Q/U$ and $V$ after than those before the cross-talk removal for both method I and II. Figure \ref{profile_plot_beforeaftercorr_method12} shows the spectra of $V$ and $Q$/$U$ before and after correction by method I (top row) and II (bottom row) labelled for the point indicated in Figure \ref{iquvpdvelo_local} by the black square 2, where the dotted lines are the observed spectrum and the solid ones are the corrected spectrum after crosstalk removal. It can be found that the modifications are evident for this point, the aymmetry properties of $Q$ and $U$ after corrections become better than those before corrections, which means that the effects of profile of $V$ in original profile of $Q$ and $U$ are significantly weakened. 
While the Figure \ref{profile_plot_slit_beforeaftercorr_method12} shows the spectra of a slit labelled in Figure \ref{iquvpdvelo_local} by the white dotted lines of $Q$ and $U$ before and after the corrections by methodI and II, and the correlations of images after corrections by methodI and II are printed in panels (d) and (h), It is found that the modification of pattern of $Q$ and $U$ are evident, and the correlations between $Q$/$U$ after correction by methodI and II exceed 90\% that means the consistency of methodI and II for this slits selected.

\subsection{Error Analysis in Method I}
For method I, it is under assumption that $V$ is known, namely $V_{obs}$ $\approx$ $V_{real}$, which is not the case in the actual observations. Additionally, the restrictions of  $\delta\lambda_{1}=\delta\lambda_{2}$ should be satisfied.
However, in the actual observation the spectrum of $Q$, $U$ and $V$ can not keep their perfect and ideal symmetry properties, which inevitably lead to errors in the calculation of the cross-talk coefficients.
Hence, numerically simulate the correcting errors originated from the deviation of symmetry properties of spectral lines due to some uncertain and unknown factors are estimated in the followings.
Here, we employ an analytical solutions of the radiative transfer equation for polarized light under the Milne-Eddington (M-E) atmosphere model (\citeauthor{1956PASJ....8..108U}, \citeyear{1956PASJ....8..108U}), which can be analytically solved to obtain the Stokes $Q$, $U$, and $V$ profiles. Then, $V$ to $Q$ and $U$ polarization crosstalk is simulated by adding a certain proportion of the $V$ signal onto the $Q$ and $U$ signals, at last the added proportion $C_{q}$/$C_{u}$ is compared to those calculated $C_{q}^{cal.}$/$C_{u}^{cal.}$ by  method I.

\begin{figure}
   \centerline{\includegraphics[width=1\textwidth,clip=]{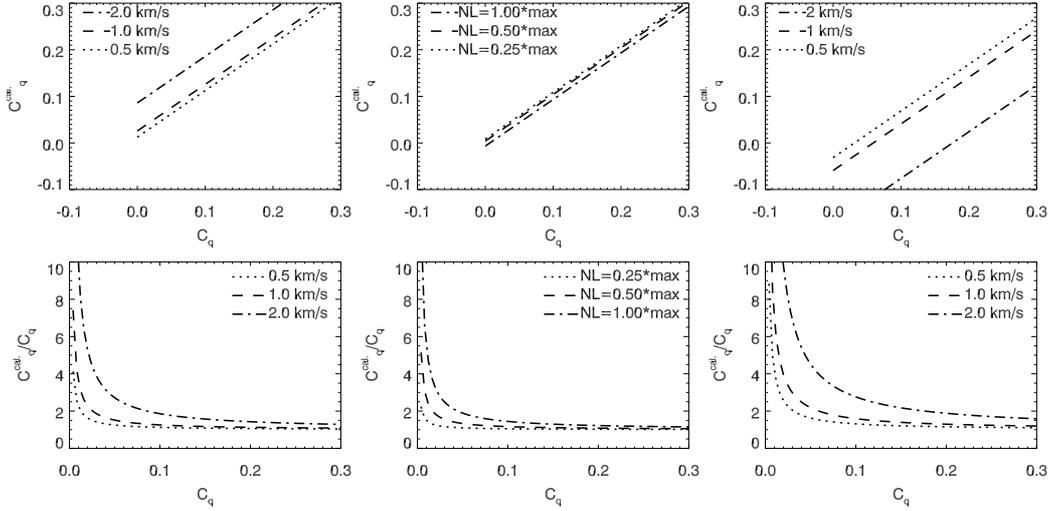}}
   \caption{The deviations between set crosstalk coefficients and the calculated ones from Method  I, here the ratio of calculated $C^{cal.}_{q}$ and $Cq$ (bottom row) and the the absolute calculated $C^{cal.}_{q}$ (up row) set are shown in Y-axis, and X-axis give the various proportion of $V$ added to $Q$ ($Cq$) that indicate the set crosstalk coefficients. The deviations due to the Doppler velocities, random noise and the combinations of two different magnetic components correspond to the left, middle and right column.} \label{cqu_Dopplernoise2comshift}
\end{figure}

In the simulations, we adjust thermodynamic parameters and compare the observation of solar spectrum ((bass2000.obspm.fr/solarspect.php), and the fix the thermodynamic parameters that meet the actual observation, while here the magnetic field related parameters are magnetic field strength $B=1000$, inclination $\gamma$=45$^{\circ}$, azimuth $\phi$=67.5$^{\circ}$. The Doppler velocity effects to calculated cross-talk are one of the main error source, which is the first aspect considered here; Random noises are another errors source should be taken into account; Another source that can create the deviations from the symetry properties are the combinations of two different magnetic components. Figure \ref{cqu_Dopplernoise2comshift} shows the deviations between the $C_{q}$ set and $C_{q}^{cal.}$ calculated by methodI, results from the Doppler velocities (left column), random noise (middle column) and the combinations of two different magnetic components (right column). For Doppler velocities effects, we add the v=0.5, 1.0 and 2.0 km/s to spectrum of $V$, $Q$ and $U$, in Figure \ref{cqu_Dopplernoise2comshift} we plot the absolute differences between the set $C_{q}$ and $C_{q}^{cal.}$ and their ratio ($C_{q}^{cal.}$/$C_{q}$) to see the relative error. It can be found that the errors become larger as the increase of velocity, the absolute errors independ of the set $C_{q}$, as a result the relative errors become small as the increase of  $C_{q}$ set. Hence, when the set $C_{q}$ is small the relative is very large, for example v=2.0 km/s the relative errors increases rapidly when the set $C_{q}$ tend to the direction of small values beyond the threshold of 0.03.

For the random noise, a certain proportion of their individual maximum works as a factor (0.25/0.50/1.00*max($Q/U/V$)), then this factors are multiplied randomly by a random number (0-1) add to individual spectrum of $V$, $Q$ and $U$ (here the number of noise is the same as that contained in one spectra of $QUV$, for the noise added at a fixed wavelength position it is created by the factor multiplying a random number). It is found that the effects of random noise are not too serious to cross-talk coefficients calculated, especially when the noise level is less than 0.5*max($Q/U/V$), which can be seen both from the the absolute difference or ratio between $C_{q}$ set and $C_{q}^{cal.}$. Even given random error of the maximum of individual Stockes parameter, the relative deviation of the cross-talk coefficient is negligible when the set coefficient is more than 0.1. 

For combinations, we use two different components of magnetic field in analytical solutions to generate and combine spectrum of $V$, $Q$ and $U$. Here magnetic field strength are set 1000 G and 2000 G with the same  inclination $\gamma$=45$^{\circ}$ and azimuth $\phi$=67.5$^{\circ}$, respectively. Another restriction needs to be added to combine these two magnetic field components is that there should exist the relative motion in longitudinal direction, which should be a condition that meets the physical situation due to the limitation of spatial resolution. 
The right column in Figure \ref{cqu_Dopplernoise2comshift} shows the results of $Cq^{cal.}$/$Cq$ and the absolute values of $Cq^{cal.}$ using spectral profiles of combining those two magnetic field components ($B$ is set 1000 G and 2000 G with the same  $\gamma$=45$^{\circ}$ and $\phi$=67.5$^{\circ}$), here the various relative velocities of 0.5, 1.0, and 2.0 km/s are shown with individual line style labelled. It is found that the deviations betweeen $Cq^{cal.}$ and $Cq$ become larger as the increase of relative velocities, the absolute errors independ of the set $C_{q}$, and as the cross-talk effect increase the relative deviation $Cq^{cal.}$ and $Cq$ tend to weak. However, when the cross-talk effects are weak, such as set cross-talk coefficients are less than 0.05, the relative deviations $Cq^{cal.}$ and $Cq$ are very large, of which the values of $Cq^{cal.}$/$Cq$ have already reached 5, although the relative errors are obvious, the absolute errors are not too serious when the set cross-talk is weak.

\subsection{Error Analysis in Method II}
\begin{figure}
   \centerline{\includegraphics[width=1\textwidth,clip=]{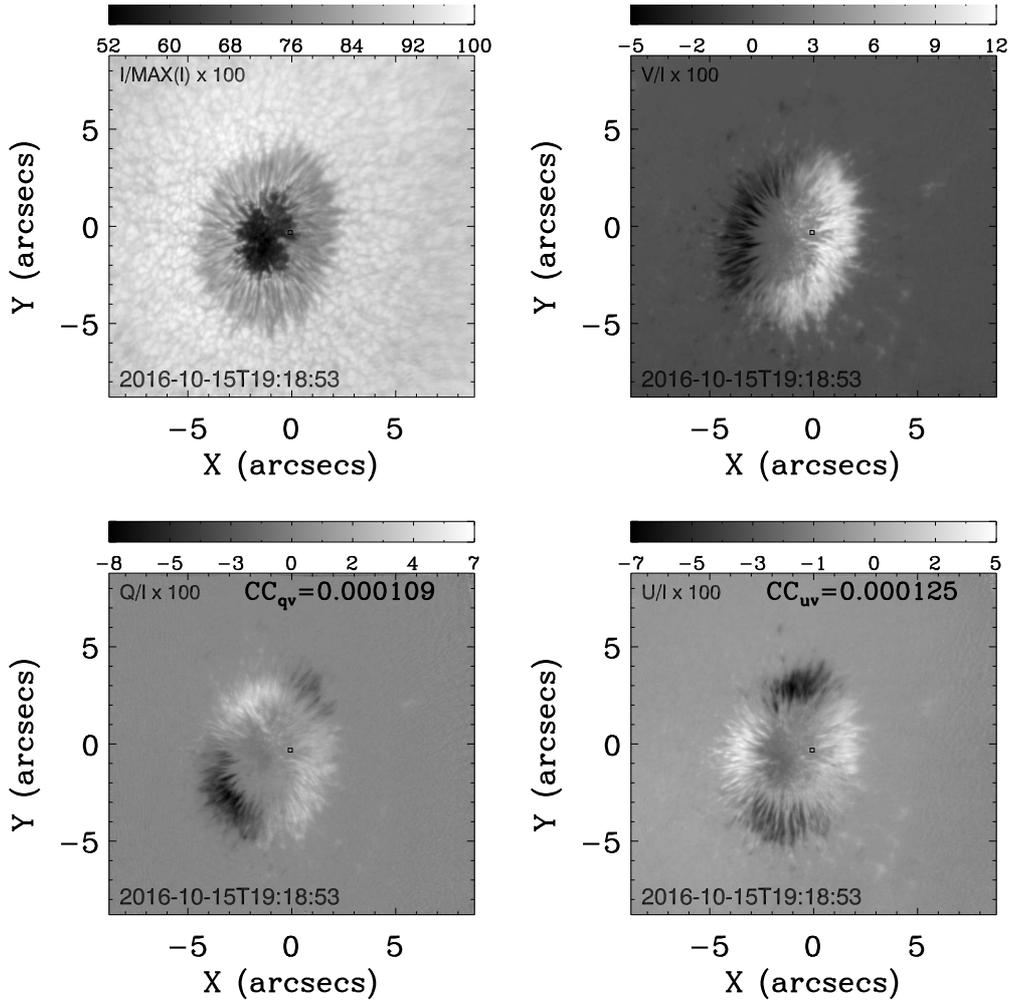}}
   \caption{The NIRS images of $I$/max($I$), $Q$/$I$, $U$/$I$, $V$/$I$ signals observed by NIRIS/GST/BBSO, and the correlations between $V$ and $Q$/$U$ are labeled by $CC_{qv}$ and $CC_{uv}$ in corresponding panels, respectively.} \label{NIRIS}
\end{figure}

\begin{figure}
 \centerline{\includegraphics[width=1\textwidth,clip=]
{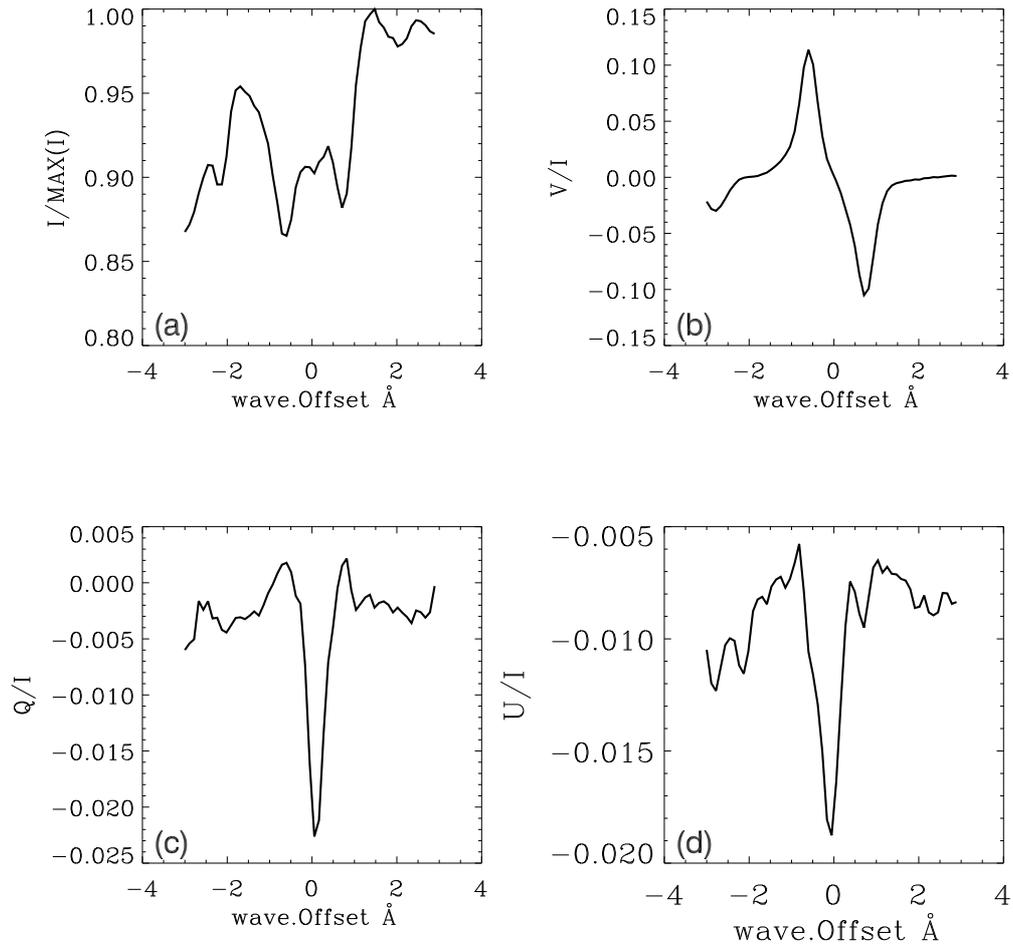}}
   \caption{The NIRIS normalized $I$/max($I$), $Q$/$I$, $U$/$I$, $V$/$I$ spectra from a position labelled by the small black squre in Figure \ref{NIRIS}.} \label{profile_plot_niris}
\end{figure}

To test the effectiveness of method II, we use a group of basical pure $I$, $Q$, $U$ and $V$ images (Figure \ref{NIRIS}) observed by NIRIS. For these data set the correlations are $1.09 \times 10^{-4}$ and $1.25 \times 10^{-4}$ for $V$ vs $Q$ and $V$ vs $U$, respectively. Because of the very small correlation, it means that there is no correlation between  $V$ and $Q$/$U$.
Next, based on set cross-talk coefficients $V$ signals are added to images of $Q$ and $U$, then method II is employed to calculate the cross-talk coefficients that can be compared with the set ones.
Table \ref{methodiiresult} shows the calculated cross-talk coefficients $Cq^{cal.}$/$Cu^{cal.}$ by method II and the corresponding set ones $C_{q}$/$C_{u}$, here the image is active region, but the same operation is done as that developed for full-disk images. Through comparisons it is found that calculated cross-talk coefficients by method II and the set ones are highly consistent without obvious deviations at all.
\begin{table}[htbp]
	\centering \caption{Comparisons of cross-talk coefficients between the set ones ($C_{q}$ and $C_{u}$) and the calculated ones by method II ($Cq^{cal.}$ and $Cu^{cal.}$).}  
	\label{methodiiresult}
	\begin{tabular}{|p{2cm}|p{2cm}|}
		\hline 
           $C_{q}$/$C_{u}$ & $Cq^{cal.}$/$Cu^{cal.}$ \\  
           \hline
            &  \\[-6pt]  
		0.2330&0.2325 \\
		0.1820&0.1825 \\
          0.1310&0.1312 \\
		0.0800&0.0800 \\
           0.0290&0.0287 \\
           -0.0220&-0.0225 \\
		-0.0730&-0.0725 \\
          -0.1240&-0.1237 \\
          -0.1750&-0.1750 \\
          -0.2260&-0.2262 \\
		\hline
	\end{tabular}
\end{table}

\begin{figure}
   \centerline{\includegraphics[width=1\textwidth,clip=]{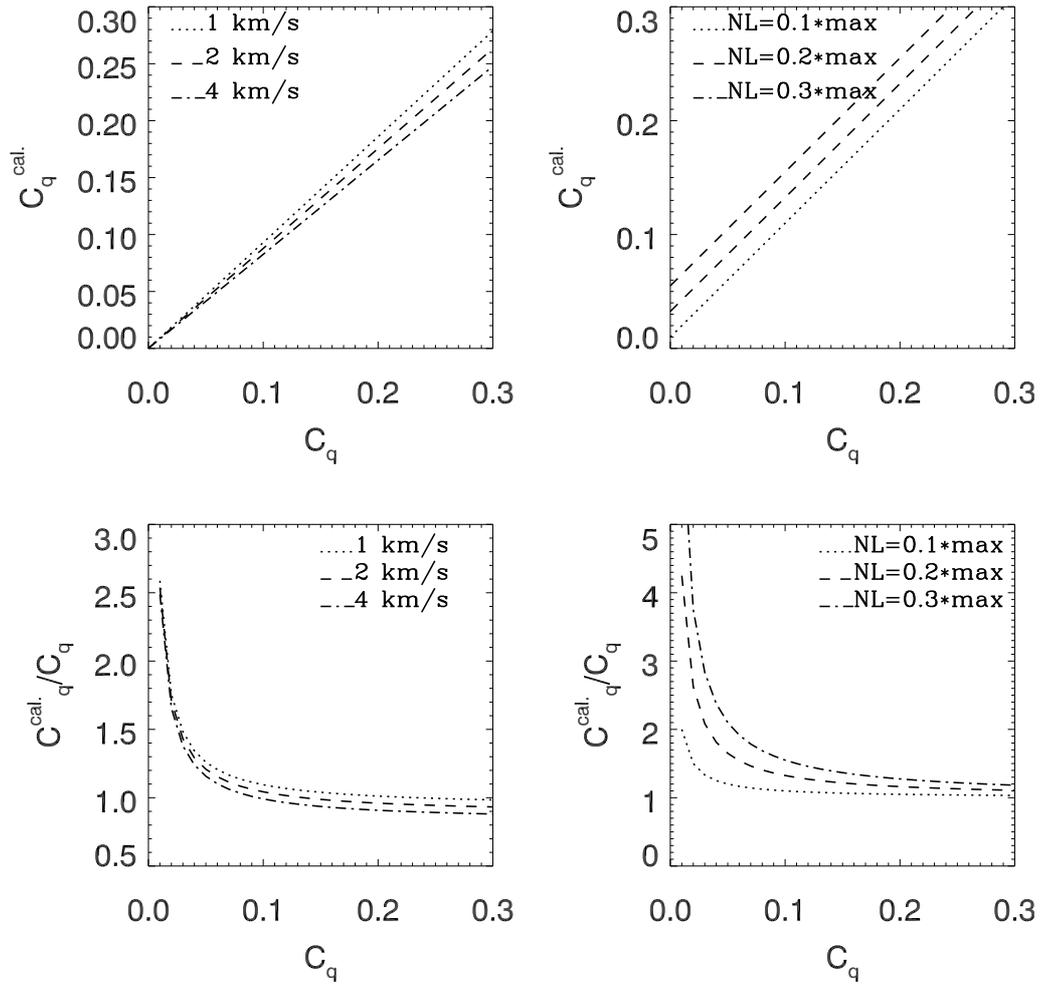}}
   \caption{The same as Figure \ref{cqu_Dopplernoise2comshift}, but for only for error sources from the Doppler velocity and random noise. Here Doppler velocity and random noise are added to observation by NIRIS (Figure \ref{NIRIS}).} \label{cqu_shiftnoise2}
\end{figure}

We tend to analyze the errors of method II based on the spectral line drift due to possibe Doppler velocity and random noise. Using the above $Q$, $U$ and $V$ images in Figure \ref{NIRIS} as the standard images (assume there is no cross-talk: see correlations in Figure \ref{NIRIS} and \ref{corrvscrossniris}). $Q$, $U$ and $V$ images with spectral position shift can be obtained by real NIRIS observations. The left column in Figure \ref{cqu_shiftnoise2} shows the error analysis due to Doppler velocities by method II, the Figure is the same as those of Figure \ref{cqu_Dopplernoise2comshift}, due to the fact that the spectral resolution of NIRIS is about 0.11 \AA, hence the Dopple velocity of 1.00, 2.00 and 4.00 km/s are considered in the calculations.
It is found that the deviations between $Cq^{cal.}$ and $Cq$ are cross-talk set dependent, which become larger as the set cross-talk increases, and the relative errors tend to be stable as the set cross-talk increase for individual Doppler velocities.
For example when the set cross-talk is bigger than 0.1, the changes in relative error almost disappeared. The right column in Figure \ref{cqu_shiftnoise2} shows the deviations between $Cq^{cal.}$ and $Cq$ vary with difference random noises level. Like Figure \ref{cqu_Dopplernoise2comshift} a set proportion of the maximum of their individual spectrum observed also work as a factor (0.1/0.2/0.3*max($Q/U/V$)). It is found that the deviations between $Cq^{cal.}$ and $Cq$ results from random noise do not depend on the values of cross-talk set like that in Figure \ref{cqu_Dopplernoise2comshift}, so when the set cross-talk become larger the relatvie errors from random noise tend to small. However, the higher level of random noise the more evident deviations can be seen. When the cross-talk larger than a threshold, the changes of relative error tends to be gentle, while the threshold is noise level dependence that threshold become larger with the increase of noise levels. For these noise level of 0.1/0.2/0.3*max($Q/U/V$), the threshold is about 0.1, which should be higher as the increase of noise levels.

\section{Discussion}\label{sec:disc}
For method I and II, it is under the assumption that the the true Stokes V is needed, but only the observed Stokes V is available, both methods should only work if V$_{obs}$ $\approx$ V$_{true}$ ($\pm$ 10 \% relative deviation). More than this for method I is that the spectrum of $Q$, $U$ and $V$ should keep their symmetry properties. The presence of asymmetries caused by gradients in velocity and magnetic field violates the main assumption of method I on the symmetry, which maybe not so relevant for the current application due to the spatial resolution with order of $>$1 arcsec/pixel.
For method II the basic idea is that Stokes $Q$, $U$ and $V$ should be independent variables, there should be low or zero correlation among $Q$, $U$ and $V$ signal. it is not important what the value of the correlation at the minimum is, but that it is a minimum. In addition the methods here should not be applied, instead of the standard method for removal of cross-talk, to calibrate polarimetric data in general, only for final clean-up of residual cross-talk. Before these applications, the better results from calibrate polarimetric data the better symmetry properties of Stokes parameters can be keeped and making V$_{obs}$ more close to V$_{true}$, so the improvement of polarimetric calibration of the telescope and instrument optics is the most important task before these applications to remove the overall cross-talk effects.

There exist evidently large cross-talk values (e.g. $>$0.4) for the SMAT data, which can be results from the regions with high level noise or without significant polarization signals. 
In application of method I to spectra without significant polarization signal is not useful. If QU and V are exhibiting only noise, the application of Eqs. (1) and (2) will only give spurious results, which will contribute to large cross-talk values of $>$40\% definitely. When we restricted to places with significant polarization signals, the large values would be reduced, however there are some large values that cannot be removed. Although the methods employed can provide a cross-talk value, but that the actual application of that cross-talk value multiplied with V$_{obs}$ instead of V$_{true}$ can worsen the cross-talk easily. For such large values, V$_{obs}$ will differ so much from V$_{true}$ that both methods cannort work anymore. The derived values of $C_{qu}$ and  the correction with -$C_{qu}*V_{obs}$ will all be off, because V$_{obs}$ deviate from V$_{true}$ very far.Additionally, 
the ratio between $C_{qu}^{true}$ and $C_{qu}^{cal.}$ can be $>>$ 1 for small cross-talk values $<$ 0.1 seen from Figure \ref{cqu_Dopplernoise2comshift}, which cases the application of the correction can actually significantly worsen the problem instead of improving it. 

These two methods have obvious correction effects in the sunspot region of strong magnetic field, which can be seee from the Figure \ref{cqcucorrectmethod1and2} in detail (e.g. region 2 marked) and magnification image of Figure \ref{cqcucorrectmethod1and2_magnif}. It seems to show that the spatial distribution of the polarization signal in $Q$ and $U$ does not resemble anymore that of $V$ after the correction, additionally there is a consistency for methodI and II to some extent (e.g. bottom row of Figure \ref{cqcucorrectmethod1and2}). It is noted that the plage regions have strong kG magnetic fields, so the magnetic fields are mainly vertical, which leads to  weak $QU$ signals. That can make $V->QU$ cross-talk stand out prominently. Although full-disk solar magnetic field is observed, we still focus on the sunspot region mainly. If the two methods in the sunspot region are effective, then they are still worth to be used in the actual observation for FMG.

Due to there is no true $Q$, $U$ and $V$ without cross-talk, even the polarimetric calibration of the telescope and instrument optics is good enough. We only do error analysis tentatively based on theories or certain assumptions. We find that once the symmetry of the spectral line is destroyed, ther errors will appear, some of which are very obvious especially for small values of cross-talk. Fortunately, it is note that although the relative errors do increase as the cross-talk coefficient decrease, the eventual effects from crosstalk are decreased because of actual weak contaminations (small values of crosstalk coefficient). While there are relatively larger crosstalk, the calculated crosstalk coefficients is more accurate which means the serious contaminations can be remove more effectively

\section{Conclusions}\label{sec:conc}
In this paper, two methods with the aim of correcting the effect of crosstalk Stokes $V$ to $Q$ and $U$ are developed and tested for full-disk magnetic field observations, which should be one of the main magnetic field data processing process for FMG/ASO. Method I extends from \citeauthor{2007ApJ...666..559S}~(\citeyear{2007ApJ...666..559S}), which is designed to deal with full-disk magnetic field observation of SMAT. Method II is a new developed one for full-disk magnetic field observation, which uses correlation information in images directly to correct the effect of crosstalk. Both methods are under the assumption of V$_{obs}$ $\approx$ V$_{true}$. 

The main ideas for method I is that the spectral profile can keep their symmetry, which can be possible due to the existence of wave shifts correcting data. Through the test that use SMAT special scanning observation designed, it is found that method I can be considered effective to correct the effects of crosstalk especially for sunspot regions. Method I would be used to process magnetic field data when FMG/ASO operates at special scanning magnetic field observation mode.

\begin{figure}
   \centerline{\includegraphics[width=1\textwidth,clip=]{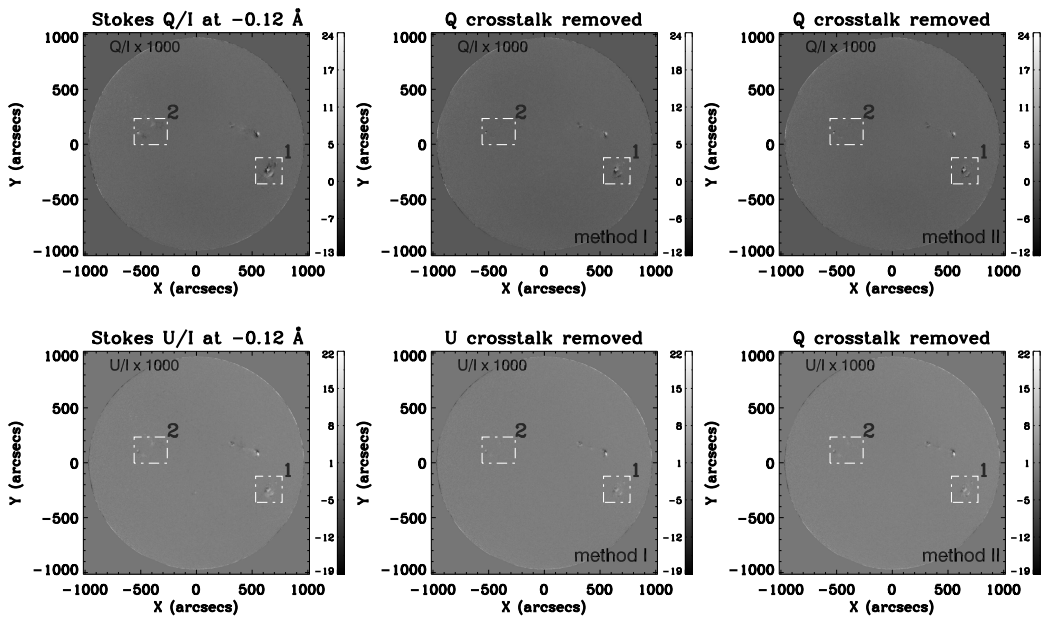}}
   \caption{{Polarization $Q$ and $U$ signals observed at -0.12\AA~ on the whole solar full-disk before and after crosstalk corrected through crosstalk coefficents of $C_{q}(x,y)$ and $C_{u}(x,y)$ in Figure \ref{cqcumethod1and2}. Two white rectangles label the regions with strong polarization signals.} \label{cqcucorrect2}} \label{cqcucorrectmethod1and2}
\end{figure}

\begin{figure}
   \centerline{\includegraphics[width=1\textwidth,clip=]{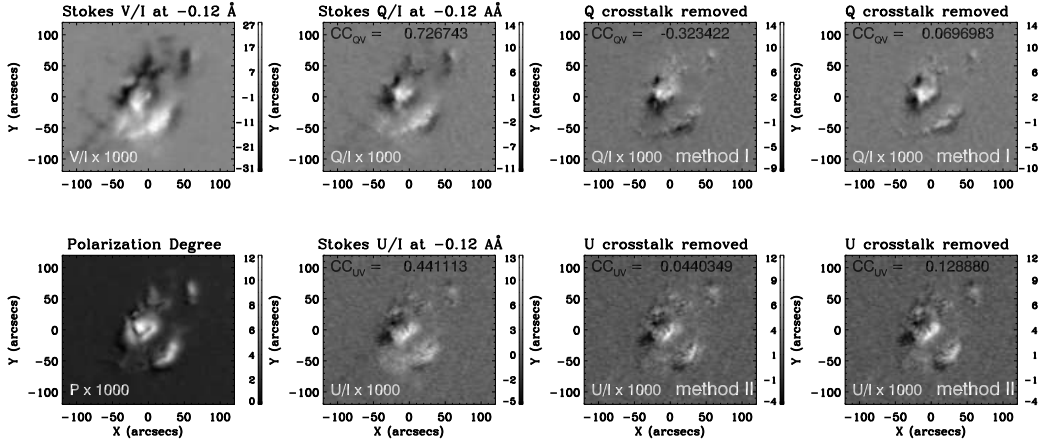}}
   \caption{The magnification images of the cut region (rectangle 1) in Figure \ref{cqcucorrectmethod1and2}, here the $V$ and polarization degree are added and the correlations of $Q$/$U$ and $V$ before and after correction are labelled in the corresponding panels by method I and II. The correlations between $V$ and $Q$/$U$ before and after correction by method I and method II are labelled in the corresponding panels.} \label{cqcucorrectmethod1and2_magnif}
\end{figure}

Method II is a novel way to correct the effect of crosstalk for full-disk magnetic field observations. The operation of method II travels on the whole full-disk using a reasonable small enough region to ensure the accuracy of local data, so this process leads to a huge consumption of computing resources. On the whole, method II can also correct the effect of crosstalk in sunspot regions that is basicall the same as method I, which can be seen from almost consistent correcting results of strong magnetic field regions in Figures \ref{cqcucorrectmethod1and2} and \ref{cqcucorrectmethod1and2_magnif}. Additionally, only one set of  $V$ to $Q$ and $U$ obtained at one position of spectral line is needed and enough for method II, which is the routine observation mode for FMG/ASO and is the evident advantage of method II. 

Both of method I and II work well at the sunspot regions, which are our main areas of concern for the full-disk observations. While the methods maybe work off at the region without significant polarization signals or the region with high level noise. However, two method I and II tested for FMG magnetic field observation can complement and authenticate each other, and the two methods can backup each other under some specific conditions.\\


This work is supported by the Strategic Priority Research Program
on Space Science, the Chinese Academy of Sciences (Grant No. XDA15320301, XDA15320302,
XDA15052200), Natural Science Foundation of China (Grant No 11203036,11703042, U1731241,11427901, 11473039, 11427901 and 11178016), the
Young Researcher Grant of National Astronomical Observations,
Chinese Academy of Sciences, and the Key Laboratory of Solar
Activity National Astronomical Observations, Chinese Academy


\end{document}